\documentclass[preprint,11pt,preprintnumbers,nofootinbib]{revtex4}
\usepackage{graphicx}
\usepackage{dcolumn}
\usepackage{bm}
\usepackage{amsfonts}

\begin{document}

\date{\today}
\title{Non-Gaussianities in N-flation}

\author{Diana Battefeld} \email[email: ]{D.Battefeld@damtp.cam.ac.uk}
\author{Thorsten Battefeld } \email[email: ]{T.Battefeld@damtp.cam.ac.uk}

\affiliation{
DAMTP, Center for Mathematical Sciences, University of Cambridge, Wilberforce
Road, Cambridge,CB3 0WA, UK}

\begin{abstract}
We compute non-Gaussianities in $\mathcal{N}$-flation, a string
motivated model of assisted inflation with quadratic, separable
potentials and masses given by  the Mar\v{c}enko-Pastur
distribution.  After estimating parameters characterizing the bi- and trispectrum in the horizon
crossing approximation, we focus on the non-linearity parameter
$f_{NL}$, a measure of the bispectrum; we compute its magnitude for narrow and broad spreads of
masses, including the evolution of modes after horizon crossing. We
identify additional contributions due to said evolution and show that
they are suppressed as long as the  fields are evolving slowly. This
renders $\mathcal{N}$-flation indistinguishable from simple
single-field models in this regime. Larger non-Gaussianities
are expected to arise for fields that start to evolve faster, and we suggest an analytic technique to estimate their contribution. However, such fast roll during inflation is not expected in $\mathcal{N}$-flation, leaving (p)re-heating as the main additional candidate for generating non-Gaussianities.

\end{abstract}
\maketitle
\newpage

\tableofcontents 

\section{Introduction}

The generation of a nearly scale invariant spectrum of primordial perturbations, as observed in the cosmic microwave background radiation (CMBR) \cite{COBE,Spergel:2006hy,Maxima,Boomerang} or large scale structure (LSS) surveys \cite{Tegmark:2001jh,Tegmark:2006az}, 
 is a decisive indicator of an inflationary epoch  in the early universe. During
inflation \cite{Guth}, an accelerated expansion of the early universe, perturbations are stretched beyond the Hubble radius 
 and re-enter
the late universe, making them observable to us, e.g. in the CMBR. 

A potential approach to discriminating
between inflationary models consist of measuring deviations from
purely Gaussian statistics; Most simple  models of inflation predict 
quasi scale invariant, nearly Gaussian, adiabatic perturbations. However, more
intricate models
such as multi-field inflationary ones (see e.g. \cite{Wands:2007bd} for a
review), might deviate from Gaussianity.
 Henceforth, a consideration of higher order correlation functions,
such as  the
bispectrum or trispectrum, could potentially shed light into the
fundamental physics responsible for generating primordial
fluctuations \cite{Verde:1999ij,Wang:1999vf}. 

A rough estimate of non-Gaussianities, as measured by the bispectrum,
is the non-linearity parameter $f_{NL}$, properly defined later in the text.
Currently, primordial non-Gaussian features in the cosmic microwave background radiation have not been
detected. For instance, the observational bound on the non-linearity
parameter obtained from the WMAP3 data set alone is $-54<f_{NL}<114$ \cite{Spergel:2006hy,
Creminelli:2005hu,Creminelli:2006rz}, and future experiments will
improve upon it  \cite{Hikage:2006fe,Komatsu:2001rj,Planck,Cooray:2006km}.
Hence, theoretical predictions for $f_{NL}$ in well motivated inflationary
models should be made before the increasingly improved  observational data is
available. 

All models of inflation predict, to a certain extent, some level of primordial
non-Gaussianity. In single-field
models, non-Gaussianities are small, usually of the order of the slow roll parameters \cite{Maldacena:2002vr}
because fluctuations freeze out once their wavelength crosses the Hubble radius.
In  multi-field models, however, the presence of multiple light degrees of freedom perpendicular to the adiabatic direction leads to the generation of isocurvature perturbations, which in turn allow the comoving curvature perturbation $\zeta$ to
evolve  after horizon crossing (HC). This leads to an additional source of potentially detectable non-Gaussianity \cite{Bartolo:2004if}.

One such multi-field model is assisted inflation, which was originally
proposed to relax fine tuning of potentials, (see e.g. \cite{Liddle:1998jc,Malik:1998gy,Kanti:1999vt,Kanti:1999ie}); it relies on $\mathcal{N}$ scalar fields, preferably
uncoupled, which assist each other in driving an inflationary
phase. Even though each individual field may not be able to generate an extended
period of inflation on its own, they can do so cooperatively. This phenomenological model is attractive, since super Planckian initial values of the fields can be avoided \cite{Liddle:1998jc,Kanti:1999vt,Kanti:1999ie} if the number of fields is large. A string motivated implementation of assisted inflation is $\mathcal{N}$-flation \footnote{See also \cite{Becker:2005sg,Ashoorioon:2006wc} for another implementation of assisted inflation within M-theory, making use of multiple M5-branes.} \cite{Dimopoulos:2005ac,Easther:2005zr,Piao:2006nm,Kim:2006te,Anber:2006xt,Gong:2006zp,Olsson:2007he}: here the many fields are identified with axions arising from some KKLT
compactification of type IIB string theory \cite{Easther:2005zr}. 

Although the existence of $\mathcal{N}$-flation via a concrete construction has not been
proven, it provides a test-bed for multi-field inflation. In \cite{Easther:2005zr} it is argued, but not proven, that an effective quadratic potential for each field without cross couplings may be attainable. Based on results of random matrix theory, it is further argued in \cite{Easther:2005zr} that the masses for the $\mathcal{N}$ fields, originally
considered identical in \cite{Dimopoulos:2005ac}, would conform to the Mar\v{c}enko-Pastur (MP) distribution \cite{Oravecz}. This \emph{spectrum} of masses is controlled by only two constants: the average mass and a parameter fixing the width/shape of the spectrum, which may be identified with the ratio of the number of axions to the total dimension of the moduli space in a given construction. The effect of this distribution
on the power-spectrum has been computed in \cite{Easther:2005zr,Gong:2006zp}, yielding a slightly redder spectral index.

Non-Gaussianities arising in multi-field inflation have been considered recently: the starting point is the three-point correlation function just after horizon crossing, which was first derived in \cite{Seery:2005gb}. Thereafter, one can evolve non-Gaussianities via the (non-linear) $\delta N$-formalism \cite{Lyth:2004gb,Lyth:2005fi} \footnote{This formalism is a crucial extension of the linear $\delta N$-formalism \cite{Starobinski,Sasaki:1995aw}, and needed to deal with higher order correlation functions.}. If the horizon-crossing (HC) approximation is used, models of assisted inflation become indistinguishable from their single-field analogs \cite{Kim:2006te}. This  is easily understood
since, by discarding the evolution of modes after horizon crossing, the main distinguishing feature of multi-field models is neglected. As a result, the spectrum of scalar and tensor perturbations is exactly the same \footnote{There is a subtlety regarding initial conditions in $\mathcal{N}$-flation, since there is no attractor solution for quadratic potentials, making $\mathcal{N}$-flation sensitive to the initial field values -- we come back to this issue later on.} to the one generated by an effective single-field model \cite{Lyth:1998xn}.

In this paper we focus on $\mathcal{N}$-flation with an arbitrary, but
large number of fields, so that the Mar\v{c}enko-Pastur distribution
can be used. We consider narrow and broad mass spectra, extending the
study of \cite{Battefeld:2006sz} where the formalism employed in this
study was developed; its application, however, was limited to simple
toy models. Here, we  analytically compute the non-linearity parameter
$f_{NL}$ without assuming the horizon crossing approximation, but within
 the slow roll approximation. A comparison with simple estimates as
well as the HC-limit, which is re-derived for completeness, reveals
that additional contributions due to the evolution of modes after
horizon crossing are present, but their magnitude is limited to a few
percent of the HC result. Hereafter, we relax  the slow roll
approximation and argue that large, but possibly transient
contributions to $f_{NL}$ should be expected from faster rolling fields,
which are however not expected during $\mathcal{N}$-flation, since heavy fields should evolve slowly up until (p)re-heating commences. Nevertheless, an analytic method to retain the main physical effect of fields starting to roll faster during inflation is suggested: an effective single-field model with steps in its potential.

The concrete outline is as follows: in Section \ref{sec:1} we review
$\mathcal{N}$-flation, focusing on the evolution during slow
roll. After that, we introduce non-Gaussianities in Section \ref{sec:ng} and provide, for later reference, the horizon crossing results of the non-linearity parameters characterizing the bi- and trispectrum, Section \ref{sec:NG_with_SR_HC}. Based on the formalism developed in \cite{Battefeld:2006sz}, we estimate $f_{NL}$ for narrow mass spectra in Section \ref{sec:narrow}, providing a simple approximation. Then, we take general broad mass spectra, properly described by the MP-distribution, to compute $f_{NL}$ without any additional approximations, Section \ref{sec:broad}. The resulting additional contributions to $f_{NL}$ due to the evolution outside the horizon are discussed and compared to previous analytic approximations, Section \ref{sec:disc}. Finally, we consider relaxing the slow roll condition in Section \ref{sec:bsr}, before we conclude in Section \ref{sec:conc}.

\section{$\mathcal{N}$-flation and Slow Roll \label{sec:1}}

We start  by considering the action for $\mathcal{N}$ scalar fields,
\begin{eqnarray}
S&=&\frac{m_p^2}{2}\int d^4x \sqrt{-g} \left(\frac{1}{2}\sum_{A=1}^{\mathcal{N}}\partial^\mu\varphi_A\partial_\mu\varphi_A+W(\varphi_1,\varphi_2,...)\right) 
\end{eqnarray}
which we assume to be responsible for driving an inflationary phase
(see e.g. \cite{Wands:2007bd} for a review on multi-field
inflation). The unperturbed volume expansion rate from an initial flat
hypersurface at $t^*$ to a final uniform density hypersurface at $t^c$
is given by  \footnote{$\mathcal{N}$ refers to the number of scalar fields,
whereas $N$ is the number of e-folds.}
\begin{eqnarray}
N(t_c,t_*)\equiv\int_*^c H dt \,,\label{nofh}
\end{eqnarray}
where $H$ is the Hubble parameter.

In $\mathcal{N}$-flation \cite{Dimopoulos:2005ac}, the $\mathcal N\sim
1000$ scalar fields that drive inflation are identified with axion fields. If one expands
the periodic axion potentials around their minima all cross-couplings
vanish \cite{Easther:2005zr}.
Hence, in the vicinity of their minima,
the fields have a potential of the form
\begin{eqnarray}
W(\varphi_1,\varphi_2,...,\varphi_{\mathcal{N}})&=&\sum_{A=1}^{\mathcal{N}}V_A(\varphi_A)\\
&=&\sum_{A=1}^{\mathcal{N}}\frac{1}{2}m_A^2\varphi_A^2\,. \label{potential}
\end{eqnarray}
We arranged the fields according to the magnitude of their masses, that is $m_A>m_B$ if $A>B$. It should be noted that $\mathcal{N}$-flation is a specific realization of assisted inflation \cite{Liddle:1998jc,Kanti:1999vt,Kanti:1999ie,Malik:1998gy} \footnote{See also \cite{Becker:2005sg} for another realization of assisted inflation based on M-theory from multiple M5-branes.}, where the many scalar fields assist each other in driving an inflationary phase, so that no single scalar field  needs to traverse a super-Planckian stretch in field space. 

\begin{figure}[tb]
  \includegraphics[scale=0.6]{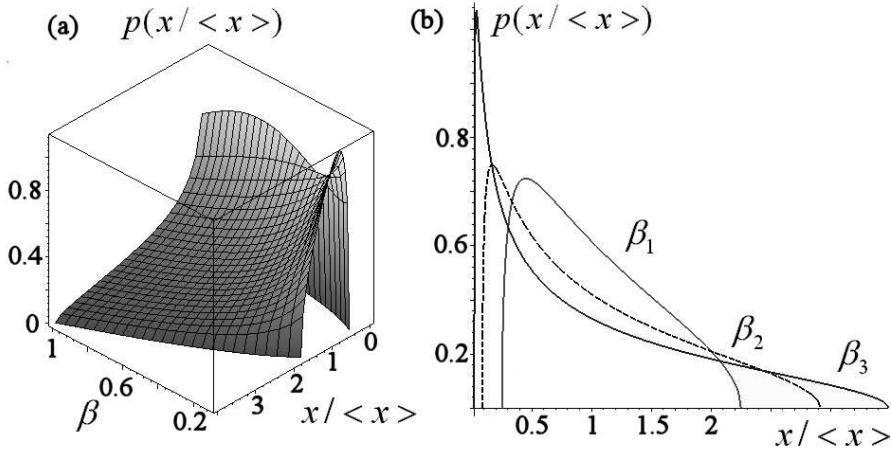}
   \caption{Probability of a given mass according to the Mar\v{c}enko-Pastur distribution from (\ref{mpdistr}), depending on $\beta$ and the dimensionless square mass $x=m^2/m_1^2$, rescaled with the expectation value $\left<x\right>$ (also dependent on $\beta$): (a) 3D-plot for $0.1<\beta<1$, (b) slices for $\beta_1=1/4$, $\beta_2=1/2$ and $\beta_3=3/4$; the closer $\beta$ is to one, the broader the mass spectrum becomes. \label{fig1}  }
\end{figure}

Further, the spectrum of masses in (\ref{potential}), which were
assumed to be equal in \cite{Dimopoulos:2005ac}, can be evaluated more
accurately by means of random matrix theory and was found by Easther and McAllister to conform to the Mar\v{c}enco-Pastur (MP) law \cite{Easther:2005zr}. This results in a probability for a given mass of 
\begin{eqnarray}
p(m^2)&=&\frac{1}{2\pi\beta m^2 \sigma^2}\sqrt{(b-m^2)(m^2-a)}\,, \label{mpdistr}
\end{eqnarray}
where $\beta$ and $\sigma$ completely describe the distribution: $\sigma$ is the average mass squared and $\beta$ controls the width and shape of the spectrum (see Fig.~\ref{fig1}). As a consequence, the smallest and largest mass are given by
\begin{eqnarray}
m_1^2=a\equiv\sigma^2(1-\sqrt{\beta})^2\,,\\
m_{\mathcal{N}}^2=b\equiv\sigma^2(1+\sqrt{\beta})^2\,.
\end{eqnarray}
In $\mathcal{N}$-flation, $\beta$ can be identified with the number of
axions contributing to inflation divided by the total dimension of the
moduli space (K\"ahler, complex structure and dilaton) in a given KKLT
compactification of type IIB string theory \cite{Easther:2005zr}. We
carry out the remainder of this analysis by treating  $\beta$ as a free parameter, but keeping in mind
that $\beta\sim 1/2$ is preferred, due to constraints arising from the
renormalization of Newton's constant \cite{Dimopoulos:2005ac}. We will
not need to specify $\sigma$, since it will cancel out; however, its
magnitude is of course constrained by the COBE normalization just as $m^2$ in chaotic inflation \cite{Gong:2006zp}. 

At this point, we introduce a convenient dimensionless mass parameter
\begin{eqnarray}
x_A&\equiv&\frac{m_A^2}{m_1^2}\,,
\end{eqnarray}
as well as the suitable short-hand notation
\begin{eqnarray}
z&\equiv&\sqrt{\beta}\,,\\
\xi&\equiv&\frac{m_{\mathcal{N}}^2}{m_1^2}=\frac{(1+z)^2}{(1-z)^2}\,.
\end{eqnarray}
Expectation values with respect to the MP-distribution can then be evaluated via
\begin{eqnarray}
\left<f(x)\right>&\equiv& \frac{1}{\mathcal{N}} \sum_{A=1}^{\mathcal{N}}f(x_i) \label{defexp0}\\
&=&\frac{(1-z)^2}{2\pi z^2}\int_1^\xi\sqrt{(\xi-x)(x-1)}\frac{f(x)}{x}\,dx\,.
\end{eqnarray}
Since  $f(x)=x^{\alpha+1}y^{\lambda x}$ will appear frequently in our
analysis, we introduce a
 more  convenient notation and define functions
$\mathcal{F}_\alpha^\lambda$, namely
\begin{eqnarray}
\mathcal{F}_\alpha^\lambda(y)\equiv\mathcal{F}_{\alpha}(y^{\lambda \xi})\,, \label{defF2}
\end{eqnarray}
where
\begin{eqnarray}
\mathcal{F}_{\alpha}(\omega)\equiv\int_{1/\xi}^1\sqrt{(1-s)(s-\xi^{-1})}s^{\alpha}\omega^s\,ds\,,
\end{eqnarray}
so that the expectation values become
\begin{eqnarray}
\left<x^{\alpha+1}y^{\lambda x}\right>=\frac{(1-z)^2}{2\pi z^2}\xi^{\alpha+2}\mathcal{F}_\alpha^{\lambda}(y)\,. \label{defexp}
\end{eqnarray}
A few useful properties of the $\mathcal{F}_\alpha$-functions and analytic approximations can be found in appendix \ref{app1}.

Throughout most of the analysis, we restrict ourselves to the slow
roll approximation. As explained above, $\mathcal{N}$ fields contribute to the
energy density of the universe through a separable potential. In this
regime, the dynamics of  $\mathcal{N}$-flation are as follows: firstly, note that the field equations and Friedman equations can be written as
\begin{eqnarray}
3H\dot{\varphi}_A &\approx&-\frac{\partial V_A}{\partial \varphi_A}\equiv- V_A^{\prime}\,,\\
3H^2&\approx& W \,.
\end{eqnarray}
Here and in the following we set the reduced Planck mass to $m_p=(8\pi G)^{-1/2}\equiv 1$. This approximation is valid if the slow roll parameters
\begin{eqnarray}
\varepsilon_A\equiv\frac{1}{2}\frac{V_{A}^{\prime 2}}{W^2}\hspace{0.5cm},\hspace{0.5cm}\eta_A\equiv\frac{V_{A}^{\prime\prime}}{W}\,, \label{srparameters}
\end{eqnarray}
are small ($\varepsilon_i\ll 1$, $\eta_i\ll 1$) and 

\begin{eqnarray}
\varepsilon\equiv\sum_{A=1}^{\mathcal N}\varepsilon_A\ll1 
\end{eqnarray}
holds. The number of e-folds of inflation becomes
\begin{eqnarray}
N(t_c,t_*)=-\int_*^c \sum_{i=1}^{\mathcal N}\frac{V_A}{V_A^\prime}d\varphi_A \,,\label{defN}
\end{eqnarray}
and the field equations can then be integrated to yield 
\begin{eqnarray}
\frac{\varphi_A^c}{\varphi_A^*}=\left(\frac{\varphi_B^c}{\varphi_B^*}\right)^{{m_A^2/m_{B}^2}}\,. \label{solphi}
\end{eqnarray}
Notice that this relationship between fields does not correspond to an
attractor solution. As a result, predictions of $\mathcal{N}$-flation can
depend on initial conditions -- admittedly, a less attractive feature  of the proposal.

There is another subtlety of $\mathcal{N}$-flation: if the mass
spectrum is broad, that is $\xi\gg 1$ corresponding to the limit
$\beta\rightarrow 1$, the heavier fields will drop out of slow roll
early,  even as  inflation continues (this is the case for the
preferred value of $\beta\sim 1/2$, corresponding to $\xi\sim
34$). The effect of these  fields on non-Gaussianities cannot be estimated properly by using the $\delta N$-formalism, which we employ in Section \ref{sec:NG_with_SR_HC} and \ref{sec:NG_no_HC}. We come back to this issue in Section \ref{sec:bsr}.

\section{Non-Gaussianities \label{sec:ng}}
Recent observations of e.g. the cosmic microwave background radiation
(CMBR) \cite{COBE,Spergel:2006hy,Maxima,Boomerang} or the large scale
structure of the Universe \cite{Tegmark:2001jh,Tegmark:2006az} made it possible to
measure two point statistics of scalar perturbations to  high
accuracy. Existing measurements are consistent with a
Gaussian spectrum, for which all odd correlation functions vanish,
while the even ones can be expressed in terms of the two point
function. For instance, the non-linearity parameter $f_{NL}$, a measure of
the three point function or bispectrum, is constraint to lie between
$-54<f_{NL}<114$ by the WMAP3 data alone \cite{Spergel:2006hy}. Future
CMBR measurements, e.g via the Planck satellite \cite{Planck},
are expected to tighten this bound up to $\mid f_{NL}\mid\sim 5$ (See
eg. \cite{Challinor:2006yh} for a review, and \cite{Fergusson:2006pr}
to relate a given primordial bispectrum to the one imprinted onto the
CMBR). Furthermore, an all sky survey of the the 21-cm background in the
frequency range from $14$ MHz to $40$ MHz with multipoles up to $10^5$
could potentially limit this parameter down to $\mid f_{NL}\mid\sim 0.01$
\cite{Cooray:2006km}; but it should be noted that this proposal is
highly optimistic. If such an observation is indeed possible, higher order correlation functions
might also be  within  reach of observation.

 It is expected that more complicated models of
the early Universe, such as $\mathcal{N}$-flation, lead to larger
contributions than simple single field models. The physical reason for this expectation is the
presence of isocurvature modes, which in turn cause evolution of
fluctuations even after crossing of the Hubble radius (referred to as horizon crossing in the following), see e.g. \cite{Rigopoulos:2005us}. Therefore, we will give a thorough examination of non-Gaussianities in models of $\mathcal{N}$-flation, only restricted by the slow roll approximation.

\subsection{Using Slow Roll and the Horizon Crossing Approximation \label{sec:NG_with_SR_HC}}
To estimate the magnitude of the three and four point functions we use
the non linear $\delta N$-formalism \cite{Lyth:2004gb,Lyth:2005fi}, which is an extension of the linear $\delta N$-formalism first proposed by Starobinsky in
\cite{Starobinski} and extended by Sasaki and Stewart
\cite{Sasaki:1995aw} among others
\cite{Seery:2005gb,Vernizzi:2006ve,Allen:2005ye}.
In this approach one
relates the perturbation of the volume expansion rate $\delta N$ to
the curvature perturbation $\zeta$,  which is possible if the initial hypersurface is flat and the final one is a uniform density hypersurface \cite{Sasaki:1995aw}. Notice that this quantity is conserved on large scales
in simple models, even beyond linear order
\cite{Lyth:2003im,Rigopoulos:2003ak} \footnote{The separate Universe
formalism developed by Rigopoulos and Shellard in
e.g. \cite{Rigopoulos:2003ak} is equivalent to the $\delta
N$-formalism.}. Given this relationship between the curvature perturbation and the volume expansion rate, one can evaluate the momentum independent pieces of non-linearity parameters, which are related to higher order correlation functions, in terms of the change in $N$ during the evolution of the Universe, see e.g. \cite{Byrnes:2006vq}. 

Following this approach, the power-spectrum $P_\zeta$ and the
bispectrum $B_\zeta$ can be computed. The ratio or $B_\zeta$ to $P_\zeta^2$ is proportional to the non-linearity parameter $f_{NL}$, modulo a momentum dependent prefactor and factors of $2\pi$. This computation was performed in \cite{Seery:2005gb,Vernizzi:2006ve,Allen:2005ye} with the result

\begin{eqnarray}
-\frac{6}{5}f_{NL}&=&\frac{r}{16}(1+f)+\frac{\sum_{A,B=1}^{\mathcal N} N_{A}N_{B}N_{AB}}{\left(\sum_{C=1}^{\mathcal N} N^2_{C}\right)^2}\,, \label{fnl}\\
&\equiv&\frac{r}{16}(1+f)-\frac{6}{5}f_{NL}^{(4)}\,,
\end{eqnarray}
where we introduced the short hand notation
\begin{eqnarray}
N_{A}&\equiv& \frac{\partial N}{\partial \varphi_A^*}\,,\\
N_{AB}&\equiv& \frac{\partial^2 N}{\partial \varphi_A^*\partial \varphi_B^*}\,,\\
\nonumber &\vdots&
\end{eqnarray}
and  we refer the interested reader to
\cite{Seery:2005gb,Vernizzi:2006ve} for details. We shall focus on the
second term in (\ref{fnl}) since the first one is known and rather
small \footnote{On geometrical grounds, we know $0\leq f\leq 5/6$
\cite{Maldacena:2002vr,Vernizzi:2006ve} ($f$ characterizes the shape
of the momentum triangle and is largest for an equilateral triangle),
while $r$ is the usual tensor:scalar ratio. The observational upper
limit on this quantity depends on the priors used in the fitting
process, but we can reliably estimate that $r/16 < 0.1$
\cite{Spergel:2006hy}. This bound will be improved by future
experiments, for example Clover \cite{Taylor:2004hh}, among others.}. If we further use the summation convention for capital indices, we arrive at \footnote{Note that $f_{NL}$ was already computed in \cite{Kim:2006te} within the horizon crossing approximation, but we repeat it here for completeness.}
\begin{eqnarray}
-\frac{6}{5}f_{NL}^{(4)}=\frac{N_A N_B N^{AB}}{(N_D N^D)^2}\,.
\end{eqnarray} 

To estimate the magnitude of the four point function, we use the momentum independent parameters $\tau_{NL}$ and $g_{NL}$ as introduced in \cite{Byrnes:2006vq} (see also \cite{Seery:2006js}),
\begin{eqnarray}
\tau_{NL}&=&\frac{N_{AB}N^{AC}N^B N_C}{(N_D N^D)^3}\,,\\
g_{NL}&=&\frac{25}{54}\frac{N_{ABC}N^A N^B N^C}{(N_D N^D)^3}\,.
\end{eqnarray}

As mentioned above, we use the horizon crossing approximation in this section in order to
compute the derivatives of the volume expansion rate; that is, we
assume that modes do not evolve once they cross the Hubble radius at
$t^*$. With this in mind, we can set $\varphi_A^c=0$ \footnote{Note that some authors prefer to set $\varphi_A^c=\varphi_A^*$ instead -- the different choices lead to the same results if modes are indeed frozen after horizon crossing.}. Since we also use the slow roll approximation, we can write (\ref{nofh}) as
\begin{eqnarray}
N(t_c,t_*)=-\int_*^c \sum_{A=1}^{\mathcal N}\frac{V_A}{V_A^\prime}d\varphi_A \,.\label{defN}
\end{eqnarray}
Confining ourselves  to $\mathcal{N}$-flation, that is, if we make use of
$V_A=m_A^2\varphi_A^2/2$, and employing equal energy initial
conditions, $m_A^2\varphi_A^{*2}=m_B^2\varphi_B^{*2}$, we arrive at
\begin{eqnarray}
N_A&=&\frac{V_A}{\sqrt{2\varepsilon_A}W}\,,\\
N_{AB}&=&\delta_{AB}\left(1-\frac{\eta_AV_A}{2\epsilon_AW}\right)\,,\\
N_{ABC}&=&-\frac{\delta_{AB}\delta_{AC}\eta_A\sqrt{2\varepsilon_A}}{2\varepsilon_A^2}\left( -\frac{V_A}{W}\eta_A+\varepsilon_A \right)\,,
\end{eqnarray}
where all potentials and slow roll parameters have to be evaluated at $t^*$, and we replaced first derivatives with respect to $\varphi_A^*$ by $V_A^\prime=\sqrt{2\varepsilon_A}W$.

It is now straightforward to evaluate the non-Gaussianity parameters to
\begin{eqnarray}
-\frac{6}{5}f_{NL}^{(4)}&=&\frac{1}{2N}\,,\label{fnlhc}\\
\tau_{NL}&=&\frac{1}{(2N)^2}\,,\\
g_{NL}&=& 0\,.
\end{eqnarray}
These parameters are leading order in the slow roll approximation, independent of the mass spectrum of $\mathcal{N}$-flation and
 indeed indistinguishable from a single-field model with quadratic
potential \cite{Byrnes:2006vq}. This is expected, since we neglected
the main feature distinguishing multi-field models from single-field
ones: the evolution of perturbations after horizon crossing due to the presence of isocurvature modes. The vanishing of $g_{NL}$ is due to the use of quadratic potentials so third derivatives of the potentials vanish.

\subsection{Beyond the Horizon Crossing Approximation \label{sec:NG_no_HC}}
We saw in the previous section that $\mathcal{N}$-flation is indistinguishable
from single-field models with respect to non-Gaussianities if the
horizon crossing approximation is used. Consequently, we must 
incorporate the discriminating feature -- the evolution of modes after
HC. To simplify matters, we make further use of the slow roll approximation, and restrict ourselves to computing $f_{NL}^{(4)}$, which is the most observationally constrained parameter.

The general expression of $f_{NL}^{(4)}$ was computed in \cite{Battefeld:2006sz} to
\begin{eqnarray}
-\frac{6}{5}f_{NL}^{(4)}=2\frac{\sum_{A=1}^{\mathcal N}\frac{u_A^2}{\varepsilon_A^*}\left(1-\eta_A^*\frac{u_A}{2\epsilon_A^*}\right)+\sum_{B,C=1}^{\mathcal N}\frac{u_Bu_C}{\varepsilon_B^*\varepsilon_C^*}\mathcal{A}_{BC}}{\left(\sum_{D=1}^{\mathcal N}\frac{u_D^2}{\varepsilon_D^*}\right)^2}\,, \label{f_NL}
\end{eqnarray}
where $u_A$ is given by
\begin{eqnarray}
u_A&\equiv&\frac{\Delta V_A}{W^*}+\frac{W^c}{W^*}\frac{\varepsilon_A^c}{\varepsilon^c}\,,\label{defuk}
\end{eqnarray}
with $\Delta V_A\equiv V_A^*-V_A^c>0$, and the symmetric $\mathcal{A}$-matrix 
\begin{eqnarray}
\mathcal{A}_{BC}&=&-\frac{W^2_c}{W_*^2}\left[\sum_{A=1}^{\mathcal N}\varepsilon_A\left(\frac{\varepsilon_C}{\varepsilon}-\delta_{CA}\right)\left(\frac{\varepsilon_B}{\varepsilon}-\delta_{BA}\right)\left(1-\frac{\eta_A}{\varepsilon}\right)\right]_c \,.  \label{defA}
\end{eqnarray}

To evaluate this expression we first compute the field values at $t_*$ and $t_c$ (see \cite{Battefeld:2006sz} for details). To do so, we need $2\mathcal{N}$ conditions,  given by the $\mathcal{N}-1$ dynamical relations between the fields from (\ref{solphi})
\begin{eqnarray}
\frac{\varphi_1^c}{\varphi_1^*}=\left(\frac{\varphi_A^c}{\varphi_A^*}\right)^{{m_1^2/m_{A}^2}}\,,\label{cond4}
\end{eqnarray}
$\mathcal{N}-1$ initial conditions, chosen for simplicity to be equal energy initial conditions\footnote{Since there is no attractor solution for quadratic potentials, there is an unavoidable dependence on initial conditions in $\mathcal{N}$-flation.}
\begin{eqnarray}
\varphi_i^*=\frac{m_1^2}{m_{\mathcal{N}}^2}\varphi_1^*\,,\label{cond3}
\end{eqnarray}
a condition that stems from (\ref{nofh}) using the requirement that $t_*$ be $N$ e-folds before $t_c$
\begin{eqnarray}
4N=\sum_{i=1}^{\mathcal{N}}\left[ (\varphi_i^{*})^2-(\varphi_i^{c})^2\right]\,,\label{cond2}
\end{eqnarray}
and the last one by demanding that slow roll ends for at least one field at
$t_c$. In the present case, the field with the largest mass leaves slow roll first \cite{Battefeld:2006sz} when
\begin{eqnarray}
\eta_{\mathcal{N}}^c=1\,. \label{cond1}
\end{eqnarray}
Once the masses are specified, solving these conditions is usually
possible. We distinguish two cases: the  first one
involves  narrow
mass spectra that have $\beta\ll 1$. This will result in a simple,
analytic expression; the second one deals with  more realistic broad mass spectra,
e.g. for $\beta\sim 1/2$. To simplify our notation, we suppress the superscript $c$ from here on. 

\subsubsection{Narrow Mass Spectra \label{sec:narrow}}
By a narrow mass spectra we mean
\begin{eqnarray}
\delta_A\equiv 1-\frac{m_1^2}{m_A^2}\ll 1\,,\\
\delta \equiv \frac{1}{\mathcal{N}}\sum_{A=1}^{\mathcal{N}}\delta_A\ll 1\,.
\end{eqnarray}
Obviously, this case corresponds to the limit $\beta\rightarrow 0$,
which is not theoretically favored, but is, nevertheless, the easiest one to consider. Using the Mar\v{c}enko-Pastur distribution, we arrive at the relation
\begin{eqnarray}
\delta&=&1-\left<x^{-1}\right>\\
&=&1-\frac{(1-z)^2}{1-z^2}\,,\label{deltaofz}
\end{eqnarray}
between $\delta$ and $z=\sqrt{\beta}$, where we made use of appendix \ref{app1} in the last step.

Using (\ref{cond1})-(\ref{deltaofz}) one can evaluate the field values $\varphi_A^*$ and $\varphi_A^c$ and the corresponding slow roll parameters, which can be found in \cite{Battefeld:2006sz}. There, it was also shown that the second term in (\ref{f_NL}) becomes
\begin{eqnarray}
\frac{\sum_{B,C=1}^{\mathcal N}\frac{u_Bu_C}{\varepsilon_B^*\varepsilon_C^*}\mathcal{A}_{BC}}{\left(\sum_{D=1}^{\mathcal N}\frac{u_D^2}{\varepsilon_D^*}\right)^2}
&=&\mathcal{O}(\delta^2/N^2)\,,\label{finalF}
\end{eqnarray}
that is, it is second order in the slow roll parameters, as well as second order in $\delta$. 

Using the same method, the first term in 
(\ref{f_NL}) becomes
\begin{eqnarray}
\frac{\sum_{A=1}^{\mathcal N}\frac{u_A^2}{\varepsilon_A^*}\left(1-\eta_A^*\frac{u_A}{2\epsilon_A^*}\right)}{\left(\sum_{D=1}^{\mathcal N}\frac{u_D^2}{\varepsilon_D^*}\right)^2}=\frac{1}{2(2N+1)}\left(1-\frac{\delta_N-\delta}{2N+1}\right)+\mathcal{O}(\delta^2)\,,
\end{eqnarray}
which includes, naturally, the contribution proportional to $1/N$, already present in the horizon crossing approximation. Hence we arrive at
\begin{eqnarray}
-\frac{6}{5}f_{NL}^{(4)}=\frac{1}{(2N+1)}\left(1-\frac{\delta_N-\delta}{2N+1}\right)+\mathcal{O}(\delta^2)\,,
\end{eqnarray}
where we only kept the leading order contribution in $\delta$. If we now use $\delta$ from (\ref{deltaofz}) and
\begin{eqnarray}
\delta_N=\frac{(1-z)^2}{(1+z)^2}\,,
\end{eqnarray}
we get after expanding in $z=\sqrt{\beta}$
\begin{eqnarray}
-\frac{6}{5}f_{NL}^{(4)}=\frac{1}{(2N+1)}\left(1-\frac{2\sqrt{\beta}}{2N+1}\right)+\mathcal{O}(\beta)\,.\label{fnlnhc}
\end{eqnarray}

This is our first major result: if we compare the above with
(\ref{fnlhc}), we observe an additional term proportional
$\sqrt{\beta}$, which vanishes if all the masses coincide. This is
expected since, in the equal mass case, there is no evolution of modes after they cross the
horizon. However,  isocurvature modes get sourced for a non-zero width of the
mass distribution. This in turn causes
modes to evolve even after their wavelength becomes larger than the
Hubble radius, leaving an imprint onto $f_{NL}^{(4)}$. Note that the
correction term in (\ref{fnlnhc}) carries an additional slow roll
suppression \footnote{See also \cite{Sasaki:2007ay} for non-Gaussianities in an exactly soluble (toy) model of
  multi-component slow-roll inflation.}. As a consequence, even if experiments improve so far as
to probe non-Gaussianities created in single-field inflationary models
(which are of order $1/N$) we will not be able to distinguish them
from $\mathcal{N}$-flation by means of their primordial non-Gaussianity, as long
as non-Gaussianities are generated during slow roll and the mass
spectrum is narrow. For simplicity, we restricted ourselves to small
$\beta$, so that expanding in terms of this small parameter is possible;
hence, the resulting correction is even more suppressed by
$\sqrt{\beta}$. Now, considering that the preferred mass spectrum in $\mathcal{N}$-flation
corresponds to $\beta\sim 1/2$, which is indeed broad, we must
compute the \emph{exact} \footnote{Exact in the statistical sense using
the large $\mathcal{N}$ limit, since we are going to use the
Mar\v{c}enko-Pastur mass distribution in order to evaluate expectation
values.} expression for $f_{NL}^{(4)}$ in the next section. Since the
resulting,  cumbersome  expression is valid even for small values of $\beta$, we can compare it with the straightforward analytic expression in (\ref{fnlnhc}) (see Fig.~\ref{fig3} and \ref{fig4}). It should be noted that the contribution due to (\ref{finalF}) will also be negligible for broad mass spectra, in agreement with the conclusions of \cite{Battefeld:2006sz}.  

\subsubsection{Broad Mass Spectra \label{sec:broad}}
In $\mathcal{N}$-flation the mass spectrum is known to conform to the
Mar\v{c}enko-Pastur distribution (\ref{mpdistr}). Armed with the
expectations values introduced in  (\ref{defexp0}), along  with
(\ref{defexp}) we can evaluate all sums
in (\ref{f_NL}).  However, before we can perform this replacement, we
need to proceed in a similar manner as we saw in the previous section, where we evaluated
the narrow mass spectrum.  Strictly speaking, we have to calculate the field values at $t^c$ and $t^*$ from (\ref{cond4})-(\ref{cond1}) in order to compute the potentials and slow roll parameters that appear in (\ref{f_NL})-(\ref{defA}).

Before we continue, we should mention another subtlety in our analysis: we take $t^c$ to be the time when the field with the largest mass leaves slow roll (\ref{cond1}); this time does not correspond to the end of inflation if the mass spectrum is stretched out considerably; since there is no cross coupling between the fields, the remaining fields can successfully continue to drive inflation even if a heavy field leaves slow roll. Consequently, several e-folds of inflation should be expected to follow after $t^c$. Henceforth, the volume expansion rate appearing in our expressions may be smaller than the usual $N\approx60$. We will discuss the effects of fields which leave slow roll while inflation continues in Section \ref{sec:bsr}.

\begin{figure}[tb]
  \includegraphics[scale=0.6]{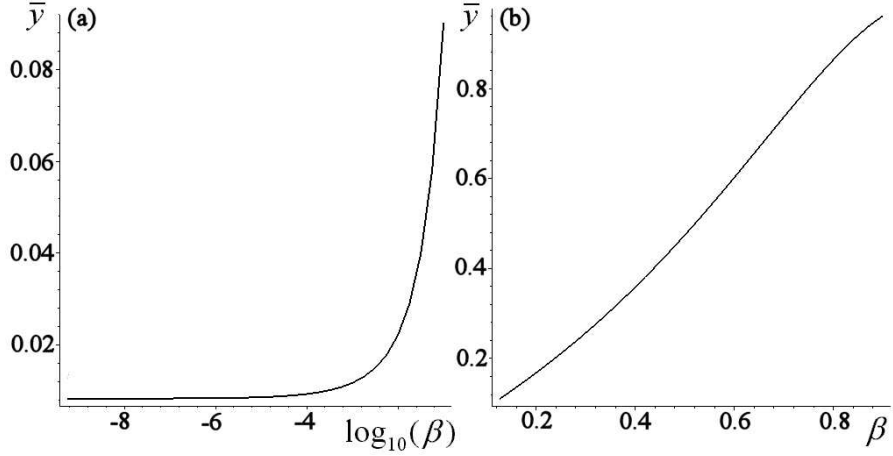}
   \caption{Solving (\ref{eqnfory2}) numerically leads to $\varphi_1^{2}/\varphi_1^{*2}\equiv\bar{y}(\beta)$ for (a) $-9\leq \log_{10}(\beta)\leq -1$, (b) $0.1 \leq \beta \leq 0.9$. We took $N=60$ in all plots. \label{fig2}  }
\end{figure}

\begin{figure}[tb]
  \includegraphics[scale=0.4]{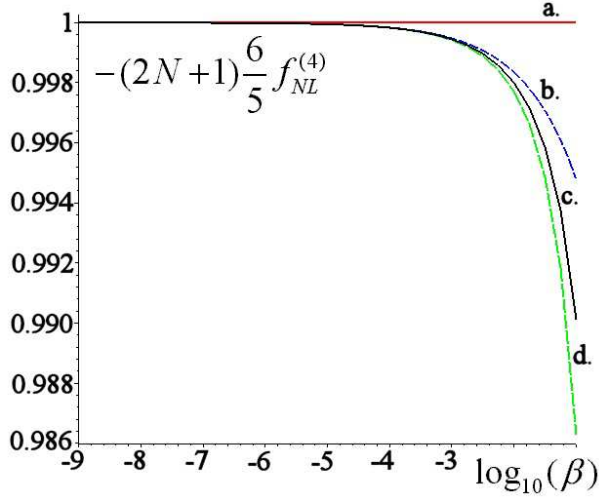}
   \caption{ $-f_{NL}^{(4)}(2N+1)6/5$ over $\log_{(10)}(\beta)$ computed using: a.~the horizon crossing approximation $-f_{NL}^{(4)}(2N+1)6/5 =1$, b.~the $\delta$-expansion from (\ref{fnlnhc}), c.~the "exact" expression from (\ref{fnlexact}) and d.~the approximation from (\ref{fapprox3}). We took $N=60$ in all plots. Note that b.~and d.~are both good approximations up until $\beta\sim 0.1$. \label{fig3}  }
\end{figure}

Given this caveat, let us now evaluate the field values $\varphi_A^*$ and $\varphi_A^c$: plugging (\ref{cond4}) and (\ref{cond3}) in (\ref{cond2}) we arrive at 
\begin{eqnarray}
\varphi_1^{*2}=\frac{4N}{\mathcal{N}}\frac{1}{\left<x^{-1}\right>-\left<x^{-1}y^x\right>} \label{phi1ofy}
\end{eqnarray}
where we replaced the sums by expectation values as introduced in (\ref{defexp0}) and defined
\begin{eqnarray}
y&\equiv&\frac{\varphi_1^{2}}{\varphi_1^{*2}}\,.
\end{eqnarray}
On the other hand, (\ref{cond1}) becomes
\begin{eqnarray}
\frac{\xi}{2N}=\frac{\left<y^x\right>}{\left<x^{-1}\right>-\left<x^{-1}y^x\right>} \label{eqnfory}\,,
\end{eqnarray}
after using (\ref{cond4}) and (\ref{cond3}) as well as (\ref{phi1ofy}). Equation (\ref{eqnfory}) is now uncoupled and needs to be solved for $y$. Using the definitions of the $\mathcal{F}$-functions in (\ref{defF2}), we can write (\ref{eqnfory}) also as
\begin{eqnarray}
0=1-2N\frac{\mathcal{F}_{-1}^{1}(y)}{\mathcal{F}_{-2}^{0}-\mathcal{F}_{-2}^{1}(y)}\,. \label{eqnfory2}
\end{eqnarray}
Unfortunately, one cannot solve (\ref{eqnfory2}) analytically, but it is easily done with standard numerical routines as implemented in e.g. MAPLE. If we denote the solution to (\ref{eqnfory2}) by $\bar{y}(\beta)$ (see Fig.~\ref{fig2} for a plot of $\bar{y}$ over $\beta$ for $N=60$), we arrive from (\ref{cond4})-(\ref{cond1}) at the desired field values  
\begin{eqnarray}
\varphi_{A}^{*2}&=&\frac{1}{x_A}\frac{4N}{a\mathcal{N}\left(\mathcal{F}_{-2}^{0}-\mathcal{F}_{-2}^1(\bar{y})\right)}\,,\\
\varphi_{A}^{2}&=&\frac{\bar{y}^{x_A}}{x_A}\frac{4N}{a\mathcal{N}\left(\mathcal{F}_{-2}^{0}-\mathcal{F}_{-2}^1(\bar{y})\right)}\,.
\end{eqnarray}
where we defined
\begin{eqnarray}
a\equiv \frac{(1-z)^2}{2\pi z^2}\,.
\end{eqnarray}
In the following, we  suppress the argument of the $\mathcal{F}$-functions, since it is always given by $\bar{y}$.

It is now straightforward to evaluate the slow roll parameters appearing in (\ref{srparameters}) to 
\begin{eqnarray}
\eta_A^*&=&x_A\frac{a}{2N}\left(\mathcal{F}_{-2}^{0}-\mathcal{F}_{-2}^1\right)\,,\\
\eta_A&=&\frac{x_A}{\xi}\,,\\
\varepsilon^*_A&=&\frac{\eta^*_A}{\mathcal{N}}\,,\\
\varepsilon_A&=&x_A\bar{y}^{x_A}\frac{2N}{\mathcal{N}\xi^2a\left(\mathcal{F}_{-2}^{0}-\mathcal{F}_{-2}^1\right)}\,,
\end{eqnarray}
and $u_A$ from (\ref{defuk}) to
\begin{eqnarray}
u_A=\frac{1}{\mathcal{N}}\left(1-\bar{y}^{x_A}+\bar{c}x_A\bar{y}^{x_A}\right)\,,
\end{eqnarray}
where we defined
\begin{eqnarray}
\bar{c}\equiv\frac{\mathcal{F}_{-2}^{0}-\mathcal{F}_{-2}^{1}}{2N\xi\mathcal{F}_0^1}\,.
\end{eqnarray}
After some more algebra, we can evaluate the two components of $f_{NL}^{(4)}$ in (\ref{f_NL}) to
\begin{eqnarray}
f(\beta)&\equiv&\frac{\sum_{A=1}^{\mathcal N}\frac{u_A^2}{\varepsilon_A^*}\left(1-\eta_A^*\frac{u_A}{2\epsilon_A^*}\right)}{\left(\sum_{D=1}^{\mathcal N}\frac{u_D^2}{\varepsilon_D^*}\right)^2}\\
 &=&\frac{\mathcal{G}}{4Na}\\
&&\nonumber\times\frac{\mathcal{F}_{-2}^{0}-\mathcal{F}_{-2}^{1}-\mathcal{F}_{-2}^{2}+\mathcal{F}_{-2}^{3}+\bar{c}\xi\left[\mathcal{F}_{-1}^{1}+2\mathcal{F}_{-1}^{2}-3\mathcal{F}_{-1}^{3}\right]+\bar{c}^2\xi^2\left[-\mathcal{F}_{0}^{2}+3\mathcal{F}_{0}^{3}\right]-\bar{c}^3\xi^3\mathcal{F}_{1}^{3}}{\left(\mathcal{F}_{-2}^{0}-2\mathcal{F}_{-2}^{1}+\mathcal{F}_{-2}^{2}-2\bar{c}\xi\left[\mathcal{F}_{-1}^{2}-\mathcal{F}_{-1}^{1}\right]+\bar{c}^2\xi^2\mathcal{F}_{0}^{2}\right)^2}\label{smallf}
\end{eqnarray}
and
\begin{eqnarray}
F(\beta)&\equiv&\frac{\sum_{B,C=1}^{\mathcal N}\frac{u_Bu_C}{\varepsilon_B^*\varepsilon_C^*}\mathcal{A}_{BC}}{\left(\sum_{D=1}^{\mathcal N}\frac{u_D^2}{\varepsilon_D^*}\right)^2} \label{bigF}\\
&=&-\frac{W^2}{W^{*2} H^2}\left(C-\frac{D}{\varepsilon}-\frac{AB^2}{\varepsilon^3}+\frac{2BE}{\varepsilon^2}-\frac{B^2}{\varepsilon}\right)
\end{eqnarray}
with
\begin{eqnarray}
\mathcal{G}&\equiv& a\left(\mathcal{F}_{-2}^{0}-\mathcal{F}_{-2}^{1}\right)\,, \label{definitionG}\\
\frac{W^2}{W^{*2}}&=&\frac{\xi^2\mathcal{G}^2}{4N^2}\,,\\
\varepsilon&=&\frac{2Na\mathcal{F}_{0}^{1}}{\mathcal{G}}\,,\\
A&\equiv&\frac{2Na}{\mathcal{G}}\mathcal{F}_{1}^{1}\,,\\
B&\equiv&\frac{4N^2a}{\xi\mathcal{G}^2}\left(\left[\mathcal{F}_{-1}^{1}-\mathcal{F}_{-1}^{2}\right]+\bar{c}\xi\mathcal{F}_{0}^{2}\right)\,,\\
C&\equiv&\frac{8N^3a}{\xi^2\mathcal{G}^3}\left(\mathcal{F}_{-2}^{1}-2\mathcal{F}_{-2}^{2}+\mathcal{F}_{-2}^{3}+2\bar{c}\xi\left[\mathcal{F}_{-1}^{2}-\mathcal{F}_{-1}^{3}\right]+\bar{c}^2\xi^2\mathcal{F}_{0}^{3}\right)\,,\\
D&\equiv&\frac{8N^3a}{\xi^2\mathcal{G}^3}\left(\left[\mathcal{F}_{-1}^{1}-2\mathcal{F}_{-1}^{2}+\mathcal{F}_{-1}^{3}\right]+2\bar{c}\xi\left[\mathcal{F}_{0}^{2}-\mathcal{F}_{0}^{3}\right]+\bar{c}^2\xi^2\mathcal{F}_{1}^{3}\right)\,,\\
E&\equiv&\frac{4N^2a}{\xi\mathcal{G}^2}\left(\mathcal{F}_{0}^{1}-\mathcal{F}_{0}^{2}+\bar{c}\xi\mathcal{F}_{1}^{2}\right)\,,\\
H&\equiv&\frac{2Na}{\mathcal{G}}\left(\mathcal{F}_{-2}^{0}-2\mathcal{F}_{-2}^{1}+\mathcal{F}_{-2}^{2}+2\bar{c}\xi\left[\mathcal{F}_{-1}^{1}-\mathcal{F}_{-1}^{2}\right]+\bar{c}^2\xi^2\mathcal{F}_{0}^{2}\right)\,,
\end{eqnarray}
so that
\begin{eqnarray}
-\frac{6}{5}f_{NL}^{(4)}(\beta)&=&2\left(f(\beta)+F(\beta)\right) \,. \label{fnlexact}
\end{eqnarray}
This is our second major result and will be discussed in the next section.

\subsubsection{Discussion \label{sec:disc}} 
\begin{figure}[tb]
  \includegraphics[scale=0.7]{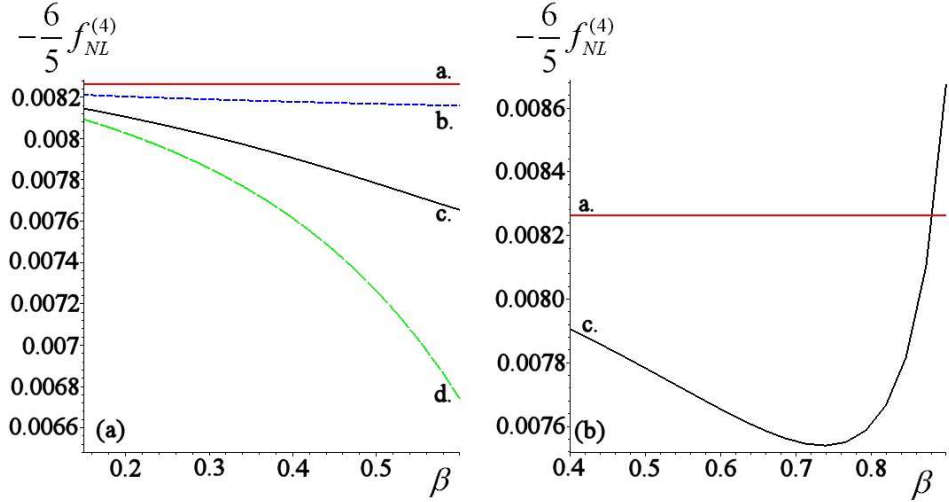}
   \caption{ $-f_{NL}^{(4)}6/5$ over $\beta$ computed using: a.~the horizon crossing approximation $-f_{NL}^{(4)}(2N+1)6/5 =1$, b.~the $\delta$-expansion from (\ref{fnlnhc}), c.~the exact expression from (\ref{fnlexact}) and d.~the approximation from (\ref{fapprox3}). We took $N=60$ in all plots. Both approximations fail to recover the turn of $f_{NL}^{(4)}$ observable in Figure (b). \label{fig4}}
\end{figure}

A plot of $f_{NL}^{(4)}$ over $\beta$ can be found in Figures
\ref{fig3} and \ref{fig4}, where it is also compared with the analytic
approximation in (\ref{fnlnhc}), the horizon crossing approximation
$-f_{NL}^{(4)}6/5=1/(2N+1)$ and the approximation in
(\ref{fapprox3}). Throughout the analysis, we took $N=60$ as a rough
estimate for the number of e-folds. It is evident that the
approximation in (\ref{fnlnhc}) is good up to $\beta\leq \bar{\beta}\sim 1/10$. In this region, the leading order contribution to the exact expression in (\ref{fnlexact}) stems from the prefactor in (\ref{smallf}), which includes a dependence on $\beta$ via $\mathcal{G}$ defined in (\ref{definitionG}), and the first summands, so that one may also use
\begin{eqnarray}
-\frac{6}{5}f_{NL}^{(4)}(\beta)&\approx&\frac{\mathcal{G}}{2Na}\frac{1}{\mathcal{F}_{-2}^{0}}\\
&=&\frac{1}{2N}\frac{\mathcal{F}_{-2}^{0}-\mathcal{F}_{-2}^{1}}{\mathcal{F}_{-2}^{0}} \label{fapprox3}
\end{eqnarray}
as an approximation for small $\beta$. Here
$\mathcal{F}_{-2}^0=(1-z)^2/(1-z^2)$ from (\ref{defF-2,0}) and
$\mathcal{F}_{-2}^{1}(\bar{y})$ is defined in (\ref{defF2}) where
$\bar{y}(\beta)$ is the solution to (\ref{eqnfory2}). Naturally, we
recover the horizon crossing result (\ref{fnlhc}) in the limit that
$\beta\rightarrow 0$.

Both, the $\delta$-expansion and the above approximation in
(\ref{fapprox3}), fail for $\beta\sim 1/2$, see
Fig.~\ref{fig4}. However, the contribution due to $F$ defined in
(\ref{bigF}) is negligible even for very broad spectra (e.g. up to
$\beta=9/10$ in Fig.~\ref{fig4} b), in agreement with the conclusions
of \cite{Battefeld:2006sz}: there, two-field models with a large ratio
of the two masses were solved analytically and an additional slow roll
suppression was found for $F$. Hence we may use 
\begin{eqnarray}
f_{NL}^{(4)}&\approx& -\frac{5}{6}\frac{\mathcal{G}}{4Na}\\
\nonumber &&\times\frac{\mathcal{F}_{-2}^{0}-\mathcal{F}_{-2}^{1}-\mathcal{F}_{-2}^{2}+\mathcal{F}_{-2}^{3}+\bar{c}\xi\left[\mathcal{F}_{-1}^{1}+2\mathcal{F}_{-1}^{2}-3\mathcal{F}_{-1}^{3}\right]+\bar{c}^2\xi^2\left[-\mathcal{F}_{0}^{2}+3\mathcal{F}_{0}^{3}\right]-\bar{c}^3\xi^3\mathcal{F}_{1}^{3}}{\left(\mathcal{F}_{-2}^{0}-2\mathcal{F}_{-2}^{1}+\mathcal{F}_{-2}^{2}-2\bar{c}\xi\left[\mathcal{F}_{-1}^{2}-\mathcal{F}_{-1}^{1}\right]+\bar{c}^2\xi^2\mathcal{F}_{0}^{2}\right)^2}
\end{eqnarray}
as an approximation in the region that is preferred in $\mathcal{N}$-flation. 

Around the preferred value of $\beta=1/2$ the magnitude of
$-f_{NL}^{(4)}$ is smaller than the horizon crossing result, but only
by a few percent (see Fig.~\ref{fig4}). Such a deviation will never
be observable. The minimum is reached for $\beta\approx 0.74$ and
$-f_{NL}^{(4)}$ increases for larger values of $\beta$ so that it
catches up with the horizon crossing result around $\beta\approx
0.88$. For even larger values of $\beta$ the magnitude of the
non-linearity parameter increases more and more, seemingly becoming significantly
large. However, one should take this result with caution, specifically the limit
$\beta\rightarrow 1$ \footnote{Remember that the limit
$\beta\rightarrow 1$ corresponds to an infinitely broad mass spectrum,
so $\beta<1$ always.}.  Here, we took the final time $t_c$ to
be the time at which the heaviest field leaves slow roll. In addition,
we assumed that sixty e-folds of inflation occurred between $t_*$
and $t_c$. However, if the spectrum of masses is indeed very broad,
there will be a considerable amount of inflation even after the
heaviest fields leave/left slow roll; hence, the potentially large
value for $-f_{NL}^{(4)}$ at $t_c$ might very well be a transient
phenomenon, due to a few heavy fields. Note that according to the
Mar\v{c}enko-Pastur distribution the majority of fields will have
relatively light masses in the broad spectrum case, see Figure
\ref{fig1} and that the majority of the masses are smaller than the
average one for $\beta$ close to one. As a result, one might
actually neglect the few heavy fields altogether, meaning, one might
want to  truncate the mass spectrum, since heavy fields will roll towards their minimum a lot faster than light ones. 

Of course, if a field leaves slow roll and starts to evolve faster, our formalism is not
applicable any more up until the field settled in its minimum. We
propose a method to estimate the production of non-Gaussianities analytically during these intervals in the next section (see also \cite{Rigopoulos:2005ae,Rigopoulos:2005us} for recent numerical work).

Before we continue, we would like to remind the reader of the
limitations of our approach: first, we focused on potentials without
any cross coupling between the fields. One can argue in favor of
vanishing couplings in the case of $\mathcal{N}$-flation if the fields
stay close to the minima of their potential \cite{Easther:2005zr}, but
in general such an assumption is rather artificial \footnote{We thank
F.~Quevedo for useful comments regarding this point. Naturally, this
means that the conditions for successful assisted inflation are hard
to satisfy.}. If such couplings are present, one should expect an
enhanced production of non-Gaussianities. We also considered only quadratic
potentials with mass spectra that conform to the Mar\v{c}enko-Pastur
distribution. We focused on this class, since the MP-distribution is
expected to properly describe the spread of masses in $\mathcal{N}$-flation in the  large $\mathcal{N}$ limit. We do not expect
qualitative differences for other spectra. Nevertheless, if the
potentials are not quadratic but quartic or exponential, we expect an
additional suppression: since an attractor solution is present for
potentials of these types (see e.g. \cite{Malik:1998gy}), isocurvature perturbations will
be suppressed and in turn any evolution of modes after horizon
crossing will also be suppressed, resulting in an additional reduction
of non-Gaussianities. Lastly, we considered equal energy initial
conditions, mainly to simplify the  computations. Since there is no attractor solution for quadratic potentials, there is a dependence on the chosen initial state. This unavoidable sensitivity to the initial configuration of fields may be considered a flaw of $\mathcal{N}$-flation, since the model becomes less predictive. However, the evident slow roll suppression of non-Gaussianities is insensitive to the chosen initial state, see e.g. the two-field cases studied in \cite{Battefeld:2006sz}.

\subsection{Beyond the Slow Roll Approximation \label{sec:bsr}}
We saw in the previous sections that multi-field inflationary models like $\mathcal{N}$-flation do not generate large non-Gaussianities during slow roll. Henceforth, one can hardly discern them from simple single-field models. We expect our conclusions to be quite general, applicable to any kind of multi-field model during slow roll, given that the potential is separable and the kinetic terms of the scalar fields are canonical. However, if fields are evolving faster, one might still produce measurable contributions to the non-linearity parameter $f_{NL}$ or higher order parameters such as $g_{NL}$ or $\tau_{NL}$.

Consider for instance the case where a few fields start to evolve faster while inflation still continues. Whenever a field behaves that way, the trajectory in field space will make a sharp turn and consequently, isocurvature perturbations will cause the adiabatic mode to evolve so that non-Gaussianities are generated. This additional source of non-Gaussianity could be recaptured by an effective one-field model: first, replace the multiple inflaton fields by one effective degree of freedom $\sigma$, which evolves due to an effective potential $V(\sigma)$ \cite{Wands:2007bd}. Doing so corresponds to the horizon crossing approximation at the perturbed level. Anyhow, this should be a good approximation since we know already that the additional contributions due to the evolution of modes after horizon crossing will hardly ever be observable. Further, assume that one of the fields leaves slow roll and starts to evolve faster. To incorporate the effect of the kink in the multi-field trajectory, introduce a step in the effective potential, followed by a subsequent re-definition of the effective model where the field under consideration is omitted. Comparing the effective model before and after removal of the field will provide the step width $\Delta \sigma$ and height $\Delta V$. Hence, we arrive at a toy model with a single inflaton field, but sharp steps in the inflaton potential. Such steps can cause considerable non-Gaussianities \cite{Chen:2006xj}, in addition to a \emph{ringing} in the power spectrum, which might actually be easier to observe than $f_{NL}$ itself \cite{Covi:2006ci,Chen:2006xj}. Subsequently, the fields that left slow roll will settle in a minimum of their potential without influencing the dynamics of the universe any more.

Nevertheless, we do not expect this case to be realized in  $\mathcal{N}$-flation: consider a broad mass spectrum
(e.g. for $\beta \geq 1/2$) as a concrete model; here, the heavy fields violate indeed the slow roll condition $|\eta_i|<1$ while inflation still goes on, but they experience an extra damping, not acceleration, since $\eta>0$. Since they evolve slower than expected from the slow roll approximation, they should not cause any additional non-Gaussianities, but simply hang around up until the light fields start to evolve faster. This will occur shortly before the onset of (p)re-heating, corresponding to the breakdown of slow roll for $\sigma$ in the effective single-field description. Consequently, we expect (p)re-heating, to be the primary additional source of non-Gaussianity (see
e.g. \cite{Barnaby:2006km}) in $\mathcal{N}$-flation. (P)re-heating \footnote{See
e.g.
\cite{Traschen:1990sw,Shtanov:1994ce,Kofman:1994rk,Kofman:1997yn,Greene:1997fu,Felder:2000hj}
for a sample on the extensive literature on (p)re-heating.} should occur in a slightly stretched-out manner
during that last one or two e-folds, since fields do not contribute all at once. Since the universe is evolving throughout, we expect expansion effects to be relevant. Henceforth, simple parametric resonance models ignoring the aforementioned expansion will not be suitable. However, stochastic resonance \cite{Kofman:1997yn}, that is, broad parametric resonance in an expanding
universe, seems to be a possibility.

A careful examination of the above mentioned topics, especially (p)re-heating, is in preparation \cite{Diana}.

\section{Conclusions \label{sec:conc}}

In this article we considered $\mathcal{N}$-flation as a concrete realization of assisted inflation motivated by string theory. Interested
in distinguishing this multi-field model from simple single-field ones, we computed primordial non-Gaussianities. 

After reviewing dynamics of $\mathcal{N}$-flation during slow roll, we evaluated non-linearity parameters characterizing the bi- and trispectrum in the horizon crossing approximation. In this limit $\mathcal{N}$-flation, and other multi-field models, become indistinguishable from their single-field analogs.

As a consequence, we incorporated the evolution of perturbations after
horizon crossing. This evolution is due to the presence of
isocurvature perturbations and provides a means of discriminating
multi-field models from single-field ones. Focusing on the non-linearity
parameter $f_{NL}$, which will be heavily constraint by observations
in the near future, we evaluated its magnitude for narrow and broad
mass distributions. In $\mathcal{N}$-flation, this mass spectrum is
described by the Mar\v{c}enko-Pastur law. We identified  additional contributions, which turned out to be only a few percent  of the horizon crossing result so that they are unobservable. 

The smallness of the additional terms is due to the slow roll
approximation employed in this paper. Henceforth, we turn our
attention to dynamics beyond slow roll;  we argue that large, but
possibly transient, contributions to $f_{NL}$ should be expected from
fast rolling fields. Such fields are not expected in case of $\mathcal{N}$-flation, but might be present in other multi field inflationary models. We suggest an effective single-field model with steps in its potential to retain the main physical effect of these fields. 

A study of (p)re-heating in $\mathcal{N}$-flation, which should provide an additional source of non-Gaussianity, is in preparation \cite{Diana}.

\begin{acknowledgments}
We would like to thank F.~Quevedo and E.~P.~S.~Shellard for useful comments and R.~Easther for many discussions and comments on the draft. D.B. would like to thank A.~C.~Davis and DAMTP at Cambridge for support. T.B. is supported by PPARC grant PP/D507366/1.

\end{acknowledgments}

\appendix

\section{The $\mathcal{F}_\alpha$ Functions \label{app1} }
Here we would like to gather some properties of the functions 
\begin{eqnarray}
\mathcal{F}_{\alpha}(\omega)\equiv\int_{1/\xi}^1\sqrt{(1-s)(s-\xi^{-1})}s^{\alpha}\omega^s\,ds\,, \label{apdefF}
\end{eqnarray}
where $\xi=(1+z)^2/(1-z)^2$ and $z$ is a free parameter \footnote{It is identified with $\sqrt{\beta}$ in $\mathcal{N}$-flation.}.
First note that analytic expressions are known if $\omega=1$: for $\alpha \geq -1$ the functions become \cite{Easther:2005zr,Oravecz} 
\begin{eqnarray}
\mathcal{F}_\alpha(1)=2\pi z^2\frac{(1-z)^2}{(1+z)^{2(\alpha+1)}}\sum_{i=1}^{\alpha+1}\frac{1}{\alpha+1}\left(\begin{array}{c}
\alpha+1\\
i
\end{array}\right)
\left(\begin{array}{c}
\alpha+1\\
i-1
\end{array}\right)z^{2(i-1)}\,,
\end{eqnarray}
by relating $\mathcal{F}$ to the moments of the Mar\v{c}enko-Pastur mass distribution, as analyzed in \cite{Oravecz}. Furthermore, the expectation values $\left<x^{-1}\right>$ and $\left<x^{-2}\right>$ were computed in \cite{Easther:2005zr}, yielding
\begin{eqnarray}
\mathcal{F}_{-2}(1)&=&\frac{(1-z)^2}{1-z^2}\frac{\xi}{a}\,, \label{defF-2,0}\\
\mathcal{F}_{-3}(1)&=&\frac{(1-z)^4}{(1-z^2)^3}\frac{\xi^2}{a}\,,
\end{eqnarray}
with $a=(1-z)^2/2\pi z^2$.

We can also write down analytic expressions for general $\omega$ in the limit $\xi\rightarrow\infty$, which corresponds to the limit $z\rightarrow 1$: first note that
\begin{eqnarray}
\bar{\mathcal{F}}_0(\omega)&\equiv&\lim_{z\rightarrow 1} \mathcal{F}_0(\omega)\\
&=&\frac{\pi \sqrt{y}}{2\ln(y)}\mathcal{I}_1(\ln(y)/2)\,,
\end{eqnarray}
where $\mathcal{I}$ is a Bessel function of the first kind. All other $\bar{\mathcal{F}}_\alpha$ can be computed via recursion since
\begin{eqnarray}
\mathcal{F}_{\alpha+1}(\omega)&=&\omega\frac{\partial  \mathcal{F}_\alpha(\omega)}{\partial\omega}\,,\\
\mathcal{F}_{\alpha-1}(\omega)&=&\int_0^{\omega} \frac{1}{\tilde{\omega}}\mathcal{F}_\alpha(\tilde{\omega}) d\tilde{\omega}\,,
\end{eqnarray}
follows directly from the definition (\ref{apdefF}).

\end{document}